\documentclass[a4paper]{article}
\usepackage[affil-it]{authblk}

\usepackage{graphicx}
\usepackage{graphics}
\usepackage{amsmath,amssymb}
\usepackage[usenames]{color}
\usepackage{subfigure}
\usepackage{hyperref}

\newcommand{\be}{\begin{equation}}
\newcommand{\ee}{\end{equation}}
\newcommand{\bea}{\begin{eqnarray}}
\newcommand{\eea}{\end{eqnarray}}
\newcommand{\ben}{\begin{enumerate}}
\newcommand{\een}{\end{enumerate}}
\newcommand{\bit}{\begin{itemize}}
\newcommand{\eit}{\end{itemize}}

\newcommand{\la}[1]{\label{#1}}

\newcommand{\Eq}[1]{Eq.~(\ref{#1})}

\newcommand{\Sec}[1]{Sec.~\ref{#1}}

\newcommand{\Fig}[1]{Fig.~\ref{#1}}

\begin{document}

\title{A dynamic picture of energy conversion in photovoltaic devices}

\author[1,2]{Robert Alicki\thanks{E-mail: \texttt{fizra@ug.edu.pl}}}
\author[3,4]{David Gelbwaser-Klimovsky\thanks{E-mail: \texttt{dgelbi@mit.edu}}}
\author[5,6]{Alejandro Jenkins\thanks{E-mail: \texttt{alejandro.jenkins@ucr.ac.cr}}}
\author[2,7]{Elizabeth von Hauff\thanks{E-mail: \texttt{e.l.von.hauff@vu.nl}}}

\affil[1]{Institute of Theoretical Physics and Astrophysics, University of Gda\'nsk, 80-952 Gda\'nsk, Poland}
\affil[2]{Freiburg Institute for Advanced Studies (FRIAS), University of Freiburg, Albertstra\ss e 19, 79104 Freiburg, Germany}
\affil[3]{Physics of Living Systems, Department of Physics, Massachusetts Institute of Technology, Cambridge, MA 02139, USA}
\affil[4]{Department of Chemistry and Chemical Biology, Harvard University, Cambridge, MA 02138, USA}
\affil[5]{Laboratorio de F\'isica Te\'orica y Computacional, Escuela de F\'isica, Universidad de Costa Rica, 11501-2060, San Jos\'e, Costa Rica}
\affil[6]{Academia Nacional de Ciencias, 1367-2050, San Jos\'e, Costa Rica}
\affil[7]{Department of Physics and Astronomy, Vrije Universiteit Amsterdam, De Boelelaan 1081, 1081 HV Amsterdam, The Netherlands}

\date{\vspace{-5ex}}

\maketitle


\begin{abstract}
Studies of emerging photovoltaics, such as organic and perovskite solar cells, have recently shown that the separation of photo-generated charge carriers is correlated with non-thermal, coherent oscillations within the illuminated device.  We consider this experimental evidence in light of results from the theory of open quantum systems that point to the need for a self-oscillating internal capacitor, acting as a microscopic piston, to explain how an illuminated solar cell operates as an autonomous heat engine.  We propose a picture of work extraction by photovoltaic devices that supersedes the quasi-static descriptions prevalent in the literature.  Finally, we argue that such a dialogue between condensed matter physics and quantum thermodynamics may offer a guide for the design of new energy transducers.
\end{abstract}

\section{Introduction}
\la{sec:intro}

Photovoltaic (PV) research is an exciting, trans-disciplinary field that combines concepts and expertise from physics, chemistry, engineering, and material science.  Advances in PV are therefore driven by the integrated efforts of researchers with broad backgrounds and training. To deliver on its promise of clean energy on a large scale, the PV community must deal with fundamental challenges and questions that range from the physics of energy conversion to the engineering of novel solar cell architectures. In this paper we re-examine a fundamental and long-standing question in the field of PV:  How does a solar cell perform electrical work?  We aim to answer this question in a way consistent with recent experimental evidence, with the laws of thermodynamics, as well as with a dynamical picture of charge carrier separation and transport.

Traditionally, PV operation has been understood as a sequence of four basic steps:
\begin{itemize}
	\item (I) absorption of light, with the corresponding generation of an electron-hole pair (exciton),
	\item (II) separation of the exciton into free charges,
	\item (III) transport of the free charges, and
	\item (IV) collection of the charge at the contact interfaces.
\end{itemize}
The total PV power conversion efficiency depends on the efficiencies of each of these steps.  However, in a seminal paper from 1961, Shockley and Queisser showed that the maximum efficiency of a single-junction solar cell may be calculated by considering the solar cell as a heat engine and applying an argument of detailed balance to its absorption and emission of radiation, in a way that gives energy and entropy budgets consistent with the laws of thermodynamics \cite{SQ}.  With this approach, they demonstrated that the maximum limit on PV efficiency can be predicted with only the knowledge of the standard solar spectrum and the semiconductor's optical bandgap. The Shockley-Queisser (SQ) limit has therefore become an important theoretical tool for estimating the performance potential of PV materials.

Nonetheless, the thermodynamics of devices that perform electrical work involves certain subtleties and potential confusions.  For a review of some of the debates concerning the application of the laws of thermodynamics to solar cells, see \cite{Markvart}.  The conventional description of photovoltaic energy conversion raises some fundamental questions about the dynamics of how electrical work is performed in a solar cell, or more specifically, what processes produce the photovoltage and photocurrent in the device and maintain these under constant illumination.
	
W\"urfel provided a bound on the maximum work that the solar cell can perform by showing that the difference in the electrochemical potential of the photogenerated electrons and holes (i.e., the splitting of the quasi-Fermi levels) in open-circuit conditions corresponds to the maximum free energy of the illuminated cell \cite{Wuerfel}, as the open-circuit voltage $V_{\rm oc}$ is directly proportional to the difference in electrochemical potentials of the photogenerated electons and holes: 
\be
V_{\rm oc} = (\mu_e - \mu_h) / q ~,
\la{eq:Voc}
\ee
where $\mu_e$ and $\mu_h$ are the electrochemical potentials of the electrons and holes, respectively, and $q$ the elementary charge.  

In the standard picture of the solar cell, no net current flows under ideal open-circuit conditions, and the photogenerated charges recombine radiatively.  Moving the solar cell out of open-circuit conditions by applying an external voltage (or, equivalently, by connecting the device to an external circuit with a load) results in a DC current through the circuit and a corresponding reduction in the photovoltage between the solar cell terminals. In ideal short-circuit conditions, all of the photogenerated charge carriers contribute to the DC photocurrent, no radiative recombination occurs, and the photovoltage is zero. Solar cells are operated at the maximum power point, at which the product of the photocurrent and photovoltage are maximized.

Interestingly, while the concepts of {\it efficiency} and {\it power} are very clearly conceptualized in photovoltaic literature, the concept of {\it work} is not.  Moreover, the underlying mechanisms that drive the spatial separation of photogenerated charge in the solar cell active layer (step II) to maintain the DC photovoltage and photocurrent have been debated for decades. In particular, there is a long-standing disagreement among experts regarding the role of the internal electric field at the $pn$ junction in silicon PV devices and also, analogously, of the energy offset at the donor-acceptor interface in organic photovoltaic (OPV) cells and of the electric field at the contact interfaces in thin-film PV. Some authors consider this electric field, in combination with the energy of the solar photon, to be the driving force responsible for charge separation and current flow in the illuminated cell \cite{Shah1999,Avrutin2011,Nelson}.  Others regard the electric field as only incidental to the solar cell's performance, and attribute the macroscopic flow of photocurrent to a gradient in chemical potential combined with selective contacts at the device terminals, that behave as semipermeable membranes. \cite{Wuerfel,Green2002}. 

According to W\"urfel,
\begin{quote}
something must be wrong in our physical education, if we think that a DC current can at all be driven in a closed circuit by a purely electrical potential difference.  The word potential alone should tell us that no energy can be gained by moving a charge along any closed path. \cite{Wuerfel}
\end{quote} 
This puzzle cannot be fully solved by invoking a spatial gradient of a chemical potential (in closed-circuit conditions) to explain charge separation and the flow of DC photocurrent, since no static potential can drive charges along a closed circuit.  Moreover, the laws of classical electrodynamics imply that only an electric field can do work directly on the charge carriers.  Further, a passive system acting as a semipermeable membrane, capable of sorting electrons and holes so as to couple them selectively to the solar cell's terminals, would constitute an unphysical Maxwell demon (see, e.g., Smoluchowski's classic treatment in \cite{Smoluchowski1912}).

In either the electrostatic or the chemical-potential picture, the energy of the incident photons is directly converted into work.  While this may seem plausible for a single photon generating one electron-hole pair in a semiconductor (the picture commonly considered in solid-state physics), converting many solar photons into a macroscopic photovoltage and photocurrent (i.e., generating power) would be at odds with the 2nd law of thermodynamics, which forbids direct conversion of heat at one temperature into work.  Though sunlight is not pure heat and has some net momentum that can, in principle, be used as a direct source of work ---for instance, by driving a solar sail--- this is not the case for solar cells, whose operation depends on the effective temperature of sunlight and not on its directionality. The thermodynamic picture of work as the macroscopic displacement of matter against a classical force has been largely missing from the theoretical treatments of PV devices and other electrical energy transducers.  For an elementary discussion of electrical work in thermodynamics, see \cite{Davies}.

The quasi-static description of PV energy conversion has led some authors to regard the passive load in the external circuit as necessary for the solar cell to perform work, by inducing the flow of DC current out of the device. However, an illuminated solar cell performs electrical work 
\be
W = Q V_{\rm oc} ~,
\la{eq:W}
\ee
inexhaustibly and with Carnot-bounded efficiency (see \Sec{sec:RE}), where $Q$ is the total amount of charge separated and then collected at the terminals.  This process is powered by the disequilibrium between the incident solar radiation and the photoactive material (see \Fig{fig:heatengine}) and is independent of any external load.

\begin{figure} [t]
	\begin{center}
		\includegraphics[width=0.6 \textwidth]{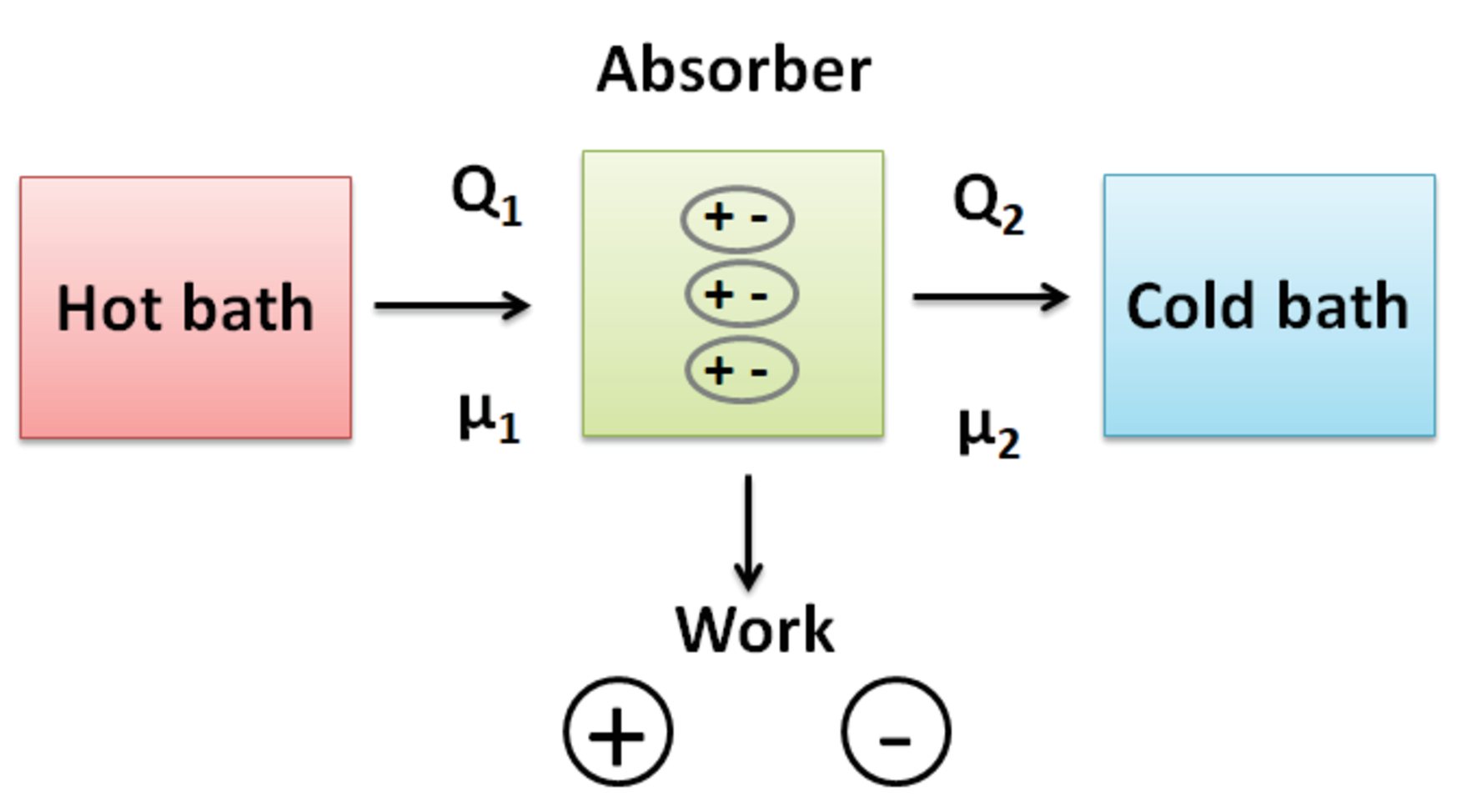}
	\end{center}
\caption{\small Schematic representation of the operation of a solar cell as a heat engine extracting work from the disequilibrium between a hot bath of solar radiation and a cold bath at ambient temperature.  This thermal disequilibrium establishes a chemical disequilibrium between electrons and holes that may be converted into electrical work by separating the photogenerated excitons.  Here we seek to clarify the dynamics that underlies this work generation.\la{fig:heatengine}}
\end{figure}

We argue that the underlying confusions about how a solar cell performs electrical work result from not taking into account the internal {\it dynamics in time} that allow the solar cell to run as an engine.  Textbook treatments of the thermodynamics of heat engines tend to neglect the role played by the cyclic mechanical motion of the engine's internal components.  In a classical engine, a piston is driven by an active, non-conservative force resulting from a positive feedback between that motion and the modulation of the coupling of the working substance to two baths that are out of equilibrium with each other. The resulting motion of the piston may be described as a {\it self-oscillation} \cite{SO}.  For a heat engine, such a dynamic description connects to the Carnot bound in the limit in which the working substance remains almost in equilibrium, so that the force on the piston, the frequency of its motion, and the power that it delivers all vanish. In practice, however, engines capable of delivering non-zero power must operate irreversibly and therefore with sub-Carnot efficiency.  On this ``finite-time'' perspective on thermodynamics, see \cite{Markvart,Ouerdane}.

A solar cell is properly understood as an engine, whose work output cannot be explained by any gradient in the quasi-static electrical or chemical potential in the active layer.  We will review the accumulating experimental evidence from emerging PVs that charge separation is associated with the non-thermal oscillation of an internal degree of freedom behaving as the engine's piston.  Recent reports of emission of THz radiation from silicon \cite{Nakanishi2012} and perovskite solar cells \cite{Guzelturk2018}, as well as IR radiation from OPV blends \cite{Falke2014}, have been identified as dynamic signatures of charge separation under open-circuit conditions.

As an analogy, consider an automobile's internal combustion engine.  One can obtain a correct upper bound on its efficiency from Carnot's theorem, without any consideration of the engine's internal dynamics.  But the engine's operation requires proper timing of the fuel's injection, ignition, and ejection with respect to the piston's cyclical motion, without which no work results from the fuel's free energy.  The relation between the quasi-static Carnot description and the actual dynamical picture in terms of a moving piston is clear enough for an automobile engine, but it has not been adequately addressed for PV or other energy transducers whose internal dynamics are microscopic and therefore difficult to observe directly.  Moreover, in the case of a gasoline engine the internal dynamics has been deliberately adjusted to optimize its performance, whereas research on improving solar cells has focused on the optical and electrical properties of semiconductors.

The first argument that there must be a self-oscillating piston within the solar cell was published in \cite{Markovian}, in the context of the Markovian master equation (MME) for the solar cell considered as an open quantum system.  A more elementary argument, directly based on the laws of thermodynamics, was subsequently published in \cite{cycle}.  In both of those works it was stressed that the coherent plasma oscillation reported in silicon semiconductors \cite{plasma1,plasma2} could serve as the required piston.  There is some experimental evidence from silicon PVs excited by laser pulses that the intensity of THz emission is correlated with the photocurrent \cite{Nakanishi2012}, but a clear verification of the connection between plasma oscillations and the separation of charge in the steadily illuminated silicon cell would require sensitive sub-millimeter spectroscopy and dedicated sample fabrication.  Recently, researchers working on hybrid perovskites reported THz emission during ultrafast charge separation under 1-sun illumination \cite{Guzelturk2018}.  The authors attributed this phenomenon to the particular properties of the perovskite lattice whose distortions produce the observed sub-millimeter radiation.  Here we will argue that oscillatory dynamics during charge separation must be common to all PV devices.

In this contribution we will pay particular attention to the evidence from OPVs.  Charge separation in OPVs is rather inefficient compared to inorganic PV technologies such as silicon, due to the large binding energy of the photogenerated excitons and the spatially localized charge carriers in molecular semiconductors. This has motivated intense research into the basic processes that govern charge separation in OPVs.  In fact, as we will discuss in \Sec{sec:pheno}, many researchers have recognized that the electrostatic picture of the OPV band diagram is insufficient to explain the performance of these solar cells. We will highlight recent breakthroughs in the understanding of charge separation in OPVs, focusing on results that underline the key role of molecular vibrations and carrier dynamics during charge separation. We shall also discuss why a dynamical picture based on molecular self-oscillations is needed to describe, in a thermodynamically consistent way, how photocharge is separated in PV devices.  We hope that this approach may open new avenues towards improving existing technologies and developing new ones. It may also help to answer longstanding, fundamental questions about the physics of emerging PVs.
	
\section{Charge separation and pumping}
\la{sec:pumping}

In \Sec{sec:intro}, we highlighted the conceptual difficulties involved in trying to account for the separation of photocharge in terms of a static electric or chemical potential.  Moreover, we stressed that the photon energy, which enters the solar cell as heat, cannot be directly converted into the work associated with this charge separation.  The solution of this paradox is to allow a time-dependent modulation of the electric field at the $pn$ junction of the PV device.  This low-entropy, coherently oscillating electric field can do work on the photogenerated charges, separating them and accelerating them ballistically towards the terminals of the solar cell, against the time-averaged electrical field pointing from anode to cathode.  This oscillating field results from the mechanical self-oscillation of an {\it internal capacitor}, powered by the thermal disequilibrium between the solar radiation and the material at room temperature.

Figure \ref{fig:capacitor} schematically represents an OPV's internal heterojunction (i.e., the donor-acceptor interface) in equilibrium. The static charge distribution at this interface may oscillate about this equilibrium position, producing the time-dependent electric field $\xi$ needed pump the charges and drive the photocurrent.  We will argue that such an oscillating field ought to be present in all PV devices and that it plays a role somewhat analogous to the oscillating electric field that injects energy into the charges circulating in a particle accelerator.  In a solar cell, the oscillation is powered by the thermal disequilibrium between the solar radiation and the photoactive material and therefore ceases if the effective temperature of the sunlight approaches the material's room temperature, in accordance with the Carnot limit.

\begin{figure} [t]
	\begin{center}
		\includegraphics[width=0.35 \textwidth]{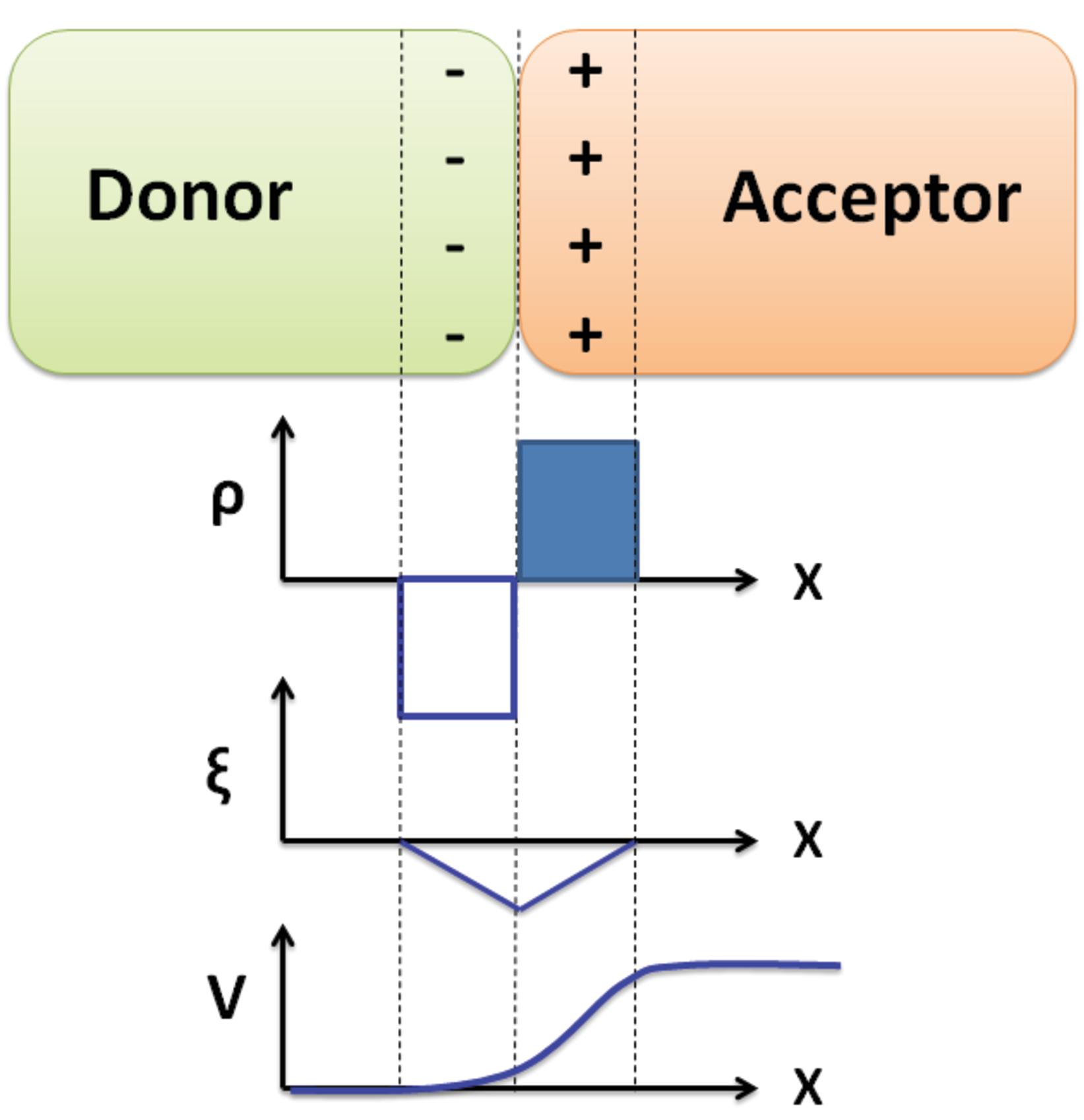}
	\end{center}
\caption{\small Physical properties of the internal capacitor at the acceptor-donor junction of an organic solar cell, in the dark and under equilibrium. Self-oscillations about the equilibrium position results in oscillations in the charge distribution $\rho$, electric field $\xi$, and electrical potential $V$ along the $x$ coordinate.\la{fig:capacitor}}
\end{figure}

To further underline the need for this internal oscillation, let us consider the photoresistor.  If the photogenerated electron and hole separate, they will diffuse through the active layer, resulting in a photocurrent.  In the presence of a net voltage difference between the two terminals, the electrons will tend to diffuse towards the anode and the holes towards the cathode.  This photocurrent discharges the terminals and dissipates electrical work, since the separation and subsequent passive diffusion of the charges produces no photovoltage.  The photoresistor therefore illustrates the fact that the generation of electric work by the illuminated solar cell requires not just the separation of the photogenerated exciton into an electron and hole to produce a photovoltage, but also an active {\it pumping} of these separated charge carriers against the net electrochemical potential difference between the two terminals.  Note that any passive selective contacts capable of generating net positive work from the separated electrons and holes would constitute Maxwell demons, as we pointed out in \Sec{sec:intro}.

Extensive theoretical and experimental research on models of nanoscale transport (a literature which thus far has made little contact with PV research) has established that pumping requires ``some sort of symmetry breaking supplemented by temporal periodicity (typically via an unbiased, nonequilibrium forcing)'' \cite{ratchet}.  The quasi-static picture of charge separation in the illuminated $pn$ junction solar cell incorporates the necessary symmetry breaking (provided by the potential step at the $pn$ interface), but not the temporal periodicity needed to describe the pumping of the separated charges.

Physically speaking, the point is that the photocurrent is pumped by an active, non-conservative force, whose power comes from an external thermodynamic disequilibrium.  In the stochastic thermodynamics approach, a non-conservative force (effectively, a negative resistance within the active device) is explicitly included in the equations of motion for the charge carriers \cite{Seifert}.  In discrete stochastic models, based on Markov chains, this non-conservative force can be modeled by ``current loops'' (also known as ``cycle fluxes'') \cite{NESS}.  Recent mathematical work has shown that the steady state resulting from such a non-conservative force may be described instead as a stochastic pump driven by an external time-dependence \cite{pumps}.  Here we argue that this time-dependence emerges dynamically, from the self-oscillation of an internal capacitor inside the solar cell.

We must also consider the question of how, and under what operating conditions, the photogenerated excitons separate into free charges within the solar cell. This question is largely ignored in the silicon PV community, in part due to the fact that the binding energy of electron-hole pairs (excitons) in silicon is around 14 meV \cite{Green2013}, which is lower than thermal energy under standard operating conditions.  The corresponding excitons may be therefore be expected to thermally ionize into free charge carriers. However, silicon solar cells also demonstrate reliable PV performance at temperatures at which thermal energy is not sufficient to ionize the excitons \cite{Loeper2012}. In emerging PV devices, such as OPVs, the process of charge separation is not as efficient and has therefore been the focus of extensive study. We explore dynamic signatures of charge separation in OPV in the following section.

\section{Phenomenology of charge separation in organic photovoltaics}
\la{sec:pheno}

An organic solar cell consists of a molecular donor-acceptor system that absorbs sunlight and separates the photogenerated excitons into free charges. Upon light absorption by the donor molecule, a tightly bound exciton is created. An electron-accepting molecule, with a lower electron affinity than the donor molecule, is used to induce electron transfer, thereby dissociating the exciton and generating photocharge. The hole is then transported along the network of donor molecules to the anode, while the electron is transported along the network of acceptor molecules to the cathode. Figure \ref{fig:DA} depicts the energy band diagram of the donor-acceptor system, including the relative energetic difference between the Highest Occupied Molecular Orbital (HOMO) and the Lowest Unoccupied Molecular Orbital (LUMO) levels of donor and acceptor molecules. After charge separation, the electron and hole may remain Coulombically bound at the donor-acceptor interface, thereby forming a localized charge transfer exciton (CTE) \cite{Deibel2010}. 

The question of the underlying mechanisms driving charge separation (exciton dissociation)) in OPV has been long debated. The ``driving energy'' or ``driving force'', $\Delta E$, refers to the energetic difference between the relative positions of the LUMO levels of the donor and acceptor molecules (Fig. \ref{fig:DA}) in the electrostatic band diagram.  This $\Delta E$ is often regarded as the excess electronic energy required to promote charge separation. This picture implies that some energy from the solar radiation is necessarily lost in the process of charge separation. A larger driving energy is generally correlated with a higher photocurrent due to efficient charge separation, but comes at the expense of a reduced open circuit voltage due to the loss in electronic energy. Simulations of OPV performance found that $\Delta E$ = 0.3 eV is optimal for maximizing photocurrent while limiting photovoltage losses \cite{Servaites2011}. Experimental evidence has shown a strong correlation between CTE energy and the open circuit voltage in a wide range of prototypical OPV devices, indicating that CTE formation is a fundamental step in charge separation \cite{Vandewal2009}. Along these lines, studies have showed that selectively exciting the donor molecule \cite{Durrant2012} or CTE \cite{Bakulin2012} with excess photon energy results in ``hot'' non-thermalized CTE states that dissociate more efficiently into free charge than thermalized ``cold'' CTE states. \cite{Durrant2012,Bakulin2012}

\begin{figure} [t]
	\begin{center}
		\includegraphics[width=0.35 \textwidth]{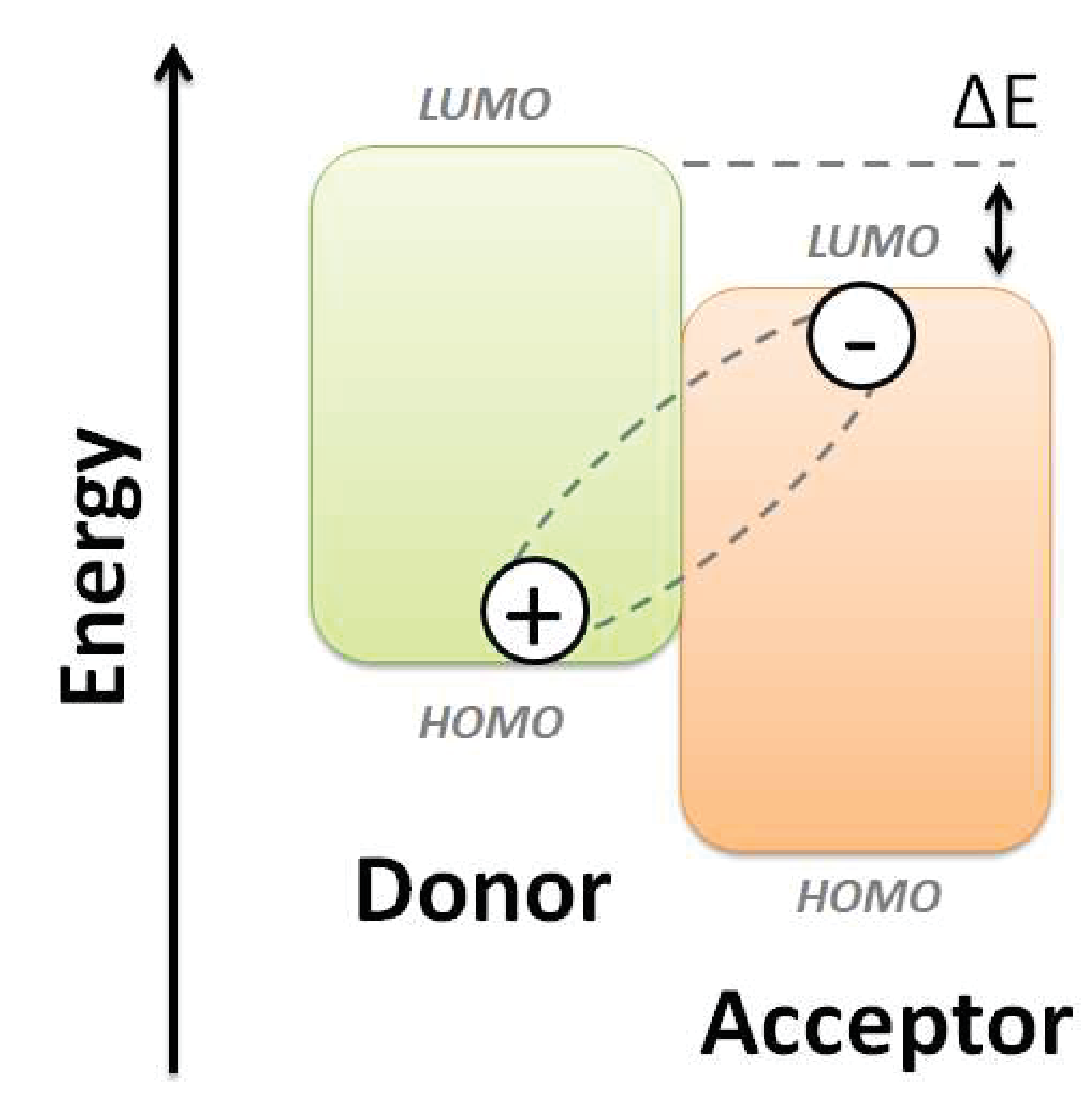}
	\end{center}
\caption{\small The energetic band diagram of an organic solar cell depicting the Highest Occupied Molecular Orbital (HOMO) and the Lowest Unoccupied Molecular Orbital (LUMO) levels of the donor and acceptor molecules. After charge separation, the electron and the hole remain Coulombically bound, forming a charge transfer exciton (CTE) at the donor-acceptor interface. The {\it driving energy} $\Delta E$ for charge separation is defined as the difference in the relative energetic position of the LUMO levels of the donor and acceptor molecules.\la{fig:DA}}
\end{figure}

However many recent reports have demonstrated that charge separation in OPV cannot be explained by the energetics of the donor-acceptor system alone, and that ultra-fast charge separation can occur in OPV systems with small  $\Delta E$ \cite{Liu2016}. In some systems, efficient charge separation has been attributed to increased delocalization of charge at the donor-acceptor interface \cite{Friend2016,Veldman2008, Hallermann2009}, leading to a larger spatial separation between the photogenerated electron and hole, and thereby a smaller Coulombic interaction. A very recent experimental study by Kurpiers et al.\ showed that charge separation in a wide range of prototypical OPV systems is both electric-field and temperature independent, with activation energies comparable to or below the thermal energy $kT$ \cite{Kurpiers2018}.  This indicates that the electrostatic picture of the donor-acceptor interface is not sufficient for understanding charge separation in OPV.

Research has been increasingly focusing on going beyond this quasi-static picture \cite{Bredas2018}, and examining the underlying dynamics of charge separation in OPV. G\'elinas et al., demonstrated that charge separation at early time-scales shows oscillatory behavior consistent with a time-dependent modulation in the Coulombic binding of the CTE via carrier delocalization \cite{Gelinas2014}. Theoretical studies have indicated that the activation energy for charge separation is reduced by increasing molecular vibrational energy \cite{Tamura2013,Tamura2008}.  Falke et al.\ reported an ultrafast electron-phonon (vibronic) coupling correlated with charge separation between a donor and acceptor molecule \cite{Falke2014}.  Bakulin et al.\ showed that selective excitation of molecular vibrational modes resulted in increased photocurrent in OPV devices \cite{Bakulin2015}.  De Sio et al.\ demonstrated how vibronic coupling leads to the ultrafast formation of free charge carriers in a pristine donor material, without the use of an acceptor molecule \cite{DeSio}. There are also interesting parallels between these observations and recent reports on natural photosynthetic systems.  Romero et al.\ demonstrated that charge separation at the photosynthetic reaction center is mediated by vibronic coupling. \cite{Romero2014}

Taken together, these results clearly point towards the importance of dynamics for understanding energy conversion in molecular systems \cite{Bredas2017}. However, a general theoretical model for the dynamics of PV energy conversion is still missing. In the next section, we will argue that the time-dependent, oscillatory phenomena reported in OPV's are consistent with what thermodynamics fundamentally requires for a dynamical description of solar cells as heat engines: that a non-equilibrium, collective, semi-classical degree of freedom must serve as a piston that extracts work and uses it to drive charge separation \cite{Markovian,cycle}. This is an important step towards closing the conceptual gap between the quasi-static picture of the solar cell associated with the SQ analysis and a realistic description of the solar cell as an engine that can provide inexhaustible electrical power under illumination.

\section{Dynamics of engine cycles}
\la{sec:RE}

Several decades of research into the mathematical description of heat engines as open quantum systems have established that work extraction by a heat engine requires a time-dependence in the Hamiltonian of the working medium, which in many cases may be interpreted as the motion of a semi-classical piston \cite{QT-review}. In the case of PV, it was noted decades ago that a thermodynamic understanding of a solar cell requires a cycle \cite{Parrott}. However it was only in \cite{Markovian, cycle} that a concrete dynamical implementation of such a cycle was proposed, based on the self-oscillation of an internal degree of freedom serving as piston.

An elementary thermodynamic argument, due independently to Rayleigh \cite{Rayleigh} and Eddington \cite{Eddington}, establishes that in any heat engine the working substance must evolve along a cycle if it is to yield net positive work.  This argument was recently extended to chemical engines and applied to solar cells in \cite{cycle}.  If the system returns to its initial thermodynamic state after a finite period, and if the system (or each of its parts) can be characterized by an instantaneous temperature \hbox{$T = \bar T + T_d$} and an instantaneous chemical potential \hbox{$\mu = \bar \mu + \mu_d$} (where $\bar T$ and $\bar \mu$ are the respective time averages), then the work performed over a full period is bounded by the first and second laws of thermodynamics as:
\be
W \leq \oint \frac{T_d}{\bar T + T_d} \cdot \delta Q + \oint \mu_d \cdot d N ~,
\la{eq:RE}
\ee
where $\delta Q$ is the heat flow into the system and $dN$ is the variation in the quantity of matter that it contains. Evidently, if $T$ and $\mu$ are constant \hbox{($T_d = \mu_d = 0$)}, then $W \leq 0$.  For a heat engine, \Eq{eq:RE} implies that efficiency is maximized by injecting heat when $T$ is maximal and rejecting heat when $T$ is minimal.  Analogously for a chemical engine, matter should be added at high chemical potential and removed at low chemical potential.  Note that a battery requires no such cyclical modulation because it's not an engine: its work output is associated with the discharging of its chemical potential, just as the work output from a hydroelectric reservoir is associated with the discharging of the water's gravitational potential.

In the quasi-static picture, the solar cell may be described as a heat engine, with the conducting electrons as the working substance, running between the hot bath of solar photons at an effective temperature $T_2$ and the cold bath of phonons at room temperature $T_1$.  In the dynamical model of \cite{Markovian}, based on the MME for the evolution of the solar cell considered as an open quantum system, the resulting open-circuit voltage is limited by
\be
q V_{\rm oc} \leq \left(1 -  \frac{T_1}{T_2} \right) E_{\rm g} ~,
\la{eq:Carnot}
\ee
where $E_{\rm g}$ is the bandgap energy of the semiconductor, while the effective photon temperature is obtained from the photon population as function of frequency, $n(\omega)$, via the Boltzmann relation:
\be
\exp \left(-\frac{\hbar \omega}{k_B T_{\rm eff}(\omega)} \right) =  \frac{n(\omega)}{n(\omega) +1} ~,
\la{eq:Teff}
\ee
so that $T_2$ in \Eq{eq:Carnot} is given by
\be
T_2 = T_{\rm eff} (E_{\rm g} / \hbar)
\la{eq:T2}
\ee
(see also \cite{localT}).  This result is consistent with the SQ analysis, as well as the experimental evidence reported in \cite{Loeper2012}.

Alternatively, one may conceptualize the illuminated solar cell as a chemical engine, with the gas of excitons within the absorber as the working substance, running between a high chemical potential $\mu_2$ and a low chemical potential $\mu_1$.  The Carnot bound in the heat-engine picture may be translated into an upper limit on this splitting $\mu_2 - \mu_1$.  Landsberg and Markvart obtained an equivalent bound in \cite{Landsberg1998}, using a quasi-static argument.  But the Rayleigh-Eddington criterion of \Eq{eq:RE} implies that any quasi-static picture of an engine must be thermodynamically incomplete.  In particular, a piston is required to extract work from the engine, by modulating the effective temperature or/and chemical potential of the working substance.

In a solar cell, the piston modulates the rate of exciton recombination in a way consistent with \Eq{eq:RE}. In the chemical engine picture, this means that fewer excitons are annihilated ($dN > 0$) at high chemical potential ($\mu_d > 0$), and more excitons are annihilated ($dN < 0$) at low chemical potential ($\mu_d < 0$).  This dynamic picture of a solar cell has important consequences for a physically realistic description of PV energy conversion, which we will explore in the following sections in the specific context of OPVs.

This chemical-engine picture can be reconciled with W\"urfel's result for the free energy of the illuminated solar cell in terms of the splitting of the quasi-Fermi levels, if we assume that a photogenerated exciton has a potential equal to $\mu_2 = \mu_e - \mu_h$ and that that same exciton is then separated by the action of a self-oscillating internal capacitor into a pair of charges with zero relative potential ($\mu_1 = 0$).  By such a process, which we will consider in more detail in \Sec{sec:dynamics}, the free energy $\mu_e - \mu_h$ can be transformed into its equivalent in electrical work $q V_{\rm oc}$, in accordance with \Eq{eq:Voc}.

\section{Internal dynamics of a solar cell}
\la{sec:dynamics}

We now reconsider the four steps in photovoltaic energy conversion enumerated in \Sec{sec:intro} by introducing a dynamical piston within the solar cell heat engine. This allows us to proceed from step I to step II, i.e. from exciton generation to charge separation, in a way consistent with the Rayleigh-Eddington criterion discussed in \Sec{sec:RE}. This piston may be described as an internal capacitor located at the donor-acceptor interface.  Self-oscillation of this capacitor about its equilibrium position is consistent with experimental reports of time-dependent charge separation in emerging PVs \cite{Gelinas2014, Falke2014}, combined with the observation of emission of radiation from the absorber layer during illumination. \cite{Guzelturk2018}

Figure \ref{fig:capacitor} qualitatively sketched the electric potential, field, and charge distribution that we expect for a solar cell's internal capacitor in equilibrium. The existence of a macroscopic equilibrium charge configuration automatically implies that collective, long-wavelength oscillations around that equilibrium are possible.  In a solid-state PV, the interface is a $pn$-junction and the charge distribution is given by essentially free charge carriers (electrons and holes).  In OPVs, the strong coupling between electronic and phononic molecular degrees of freedom implies that the charge distribution forming the internal capacitor has a more complicated structure.  A detailed ``exciton-lattice model'' of the organic heterojunction is discussed, e.g., in \cite{Bittner2014}. Polymer chains aligned along the junction support both electronic and phononic states delocalized along the interface.  Thus, the infra-red active phonon modes at the donor-acceptor interface may provide the oscillating degree of freedom for the internal capacitor at the junction. \cite{Casalegno2017}

Thermal and zero-point fluctuations of the internal capacitor alone cannot account for charge separation, as charges must overcome an overall electrostatic potential barrier in the order of 1 eV.  What is needed is a mechanism enabling power transfer from the photons into a cyclical, non-thermal motion of the piston.  This mechanism has been identified as a coherent self-oscillation associated with a positive feedback \cite{Markovian}.  In the linear regime it can be described by a set of two coupled, phenomenological mean-field type equations
\bea
\dot n = - \Gamma(x) {n} + B ~; \la{eq:n} \\
\ddot x + \gamma \dot x + \omega^2 x = A (n_0- n) ~.\la{eq:x}
\eea
in which $x$ denotes the deviation of the internal capacitor width from its equilibrium value and $n$ denotes the density of photo-generated excitons (with equilibrium value $n_0 = B/\Gamma(0)$) \cite{cycle}.  Equation (\ref{eq:n}) is a kinetic equation in which $B$ is the exciton generation rate and $\Gamma$ the recombination rate. For free oscillations of $x$, the angular frequency is $\omega$ and the damping coefficient is $\gamma$.  The term $A (n_0 - n)$ in \Eq{eq:x} corresponds to an effective pressure that acts on the piston due to the presence of a gas of photo-generated excitons in the absorber. The positive feedback required for self-oscillation is provided by the dependence of $\Gamma$ on the value of $x$ in \Eq{eq:n}, and by the dependence of the effective pressure in \Eq{eq:x} on $n$.

A linear stability analysis of Eqs.\ \eqref{eq:n} and \eqref{eq:x} (see \cite{cycle}) gives the following criterion for self-oscillation:
\be
A n_0 \Gamma'(0) < - \omega^2 \gamma ~.
\la{eq:self_cond}
\ee
This condition can be satisfied if $\Gamma'(0) < 0 $.  In such a case, the chemical disequilibrium in the illuminated solar cell can power the self-oscillation of the internal capacitor, which can in turn drive the separation of charge and the DC photocurrent in the illuminated solar cell.

Figure \ref{fig:chargeseparation} illustrates the dynamics of charge separation in the OPV cell. Expansion of the internal capacitor causes some previously photogenerated excitons to be absorbed into the region between the two plates of that capacitor, where they are subject to a strong electric force that moves electrons towards the aceptor and holes towards the donor material, respectively.  This can produce a CTE configuration like that depicted in \Fig{fig:DA}.  The same electric field within the oscillating capacitor acts as a time-dependent separating force that causes tunneling ionization of the CTE, separating the charge carriers and giving them an (initially) ballistic motion towards the corresponding terminals of the device.

Note that the subsequent contraction of the internal capacitor reduces the region where the separating force acts, thereby increasing the rate of exciton recombination.  This implies that $\Gamma'(0) < 0$, in accordance with \Eq{eq:self_cond}.  This also agrees with the Rayleigh-Eddington criterion of \Sec{sec:RE}, because the expansion of internal capacitor increases the density of the exciton gas in the absorber.  This exciton gas plays a role analogous to that of steam in a conventional steam engine. Generally, higher density implies higher chemical potential, while lower density implies lower chemical potential.

\begin{figure} [t]
	\begin{center}
		\includegraphics[width=0.7 \textwidth]{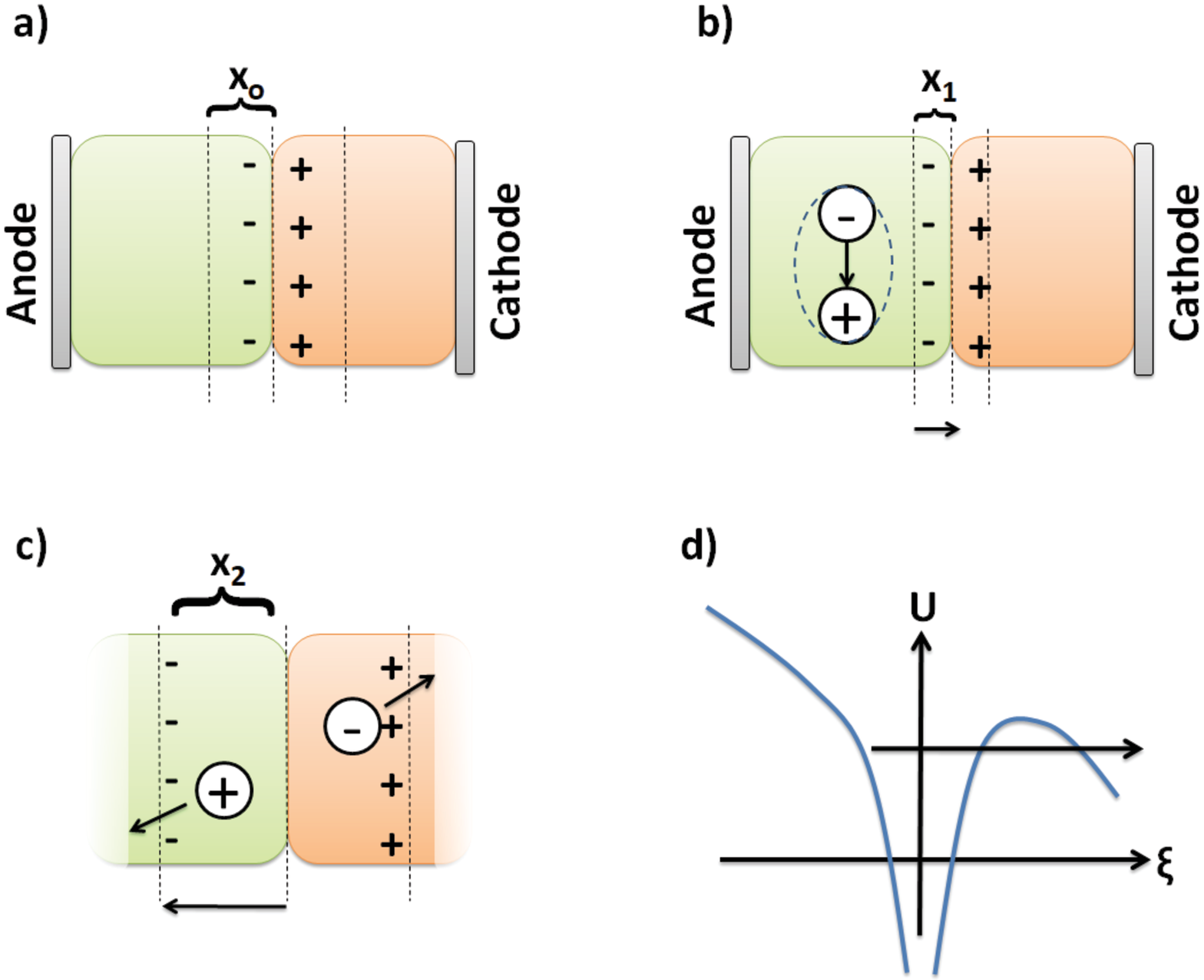}
	\end{center}
\caption{\small Schematic representation of charge separation at the donor-acceptor interface. (a) Equilibrium configuration of the internal capacitor formed by the charge bilayer around the interface, with separation $x_0$ between the two layers. (b) Contraction phase of internal capacitor oscillation, with $x_1 <  x_0$.  The electron and hole in the absorber are Coulombically bound.  (c) Expansion phase, with $x_2 > x_0$.  The strong electric force within the internal capacitor separates the exciton and ballistically accelerates the electron towards the cathode and the hole towards the anode.  (d)  Combined Coulombic and linear potential inside internal capacitor as a function of $\xi$ (the component electron-hole separation along the electric field of the internal capacitor). Charges are separated either by tunneling ionization or by direct decomposition over the potential barrier. The corresponding unstable states correspond to the CTE.\la{fig:chargeseparation}}
\end{figure}

\section{The four steps revisited}
\la{sec:steps}

Based on these results, we revisit the four steps in PV energy conversion. Step I (light absorption, and the generation of excitons) and step IV (collection of carriers at the contacts) are the same as introduced in \Sec{sec:intro}. However the progression from Step I to Step III can be now described as follows:
\begin{itemize}
	\item (II$'$-a) the piston's motion modulates the rate of exciton recombination (heat rejection) in the absorber, and
	\item (II$'$-b) the photo-generated excitons exert an effective pressure on the piston.
	\item (III$'$) The mechanical motion of the piston separates the photogenerated excitons into free charge carriers with initially ballistic motions, resulting in a photovoltage and photocurrent.
\end{itemize}
Unlike the traditional steps mentioned in \Sec{sec:intro}, the scheme I, II$'$-a, II$'$-b, III$'$, IV, is not sequential.  All four distinct processes must proceed simultaneously to account for PV energy conversion.

The action of the piston results from two distinct but interconnected processes (II$'$-a and II$'$-b), as a consequence of Eqs.\ (\ref{eq:n}) and (\ref{eq:x}), respectively. Self-oscillation of the piston occurs when the feedback is positive, thereby destabilizing the equilibrium position of the piston. This is characterized in the mathematical literature as a Hopf bifurcation \cite{SO}.  Though process III$'$ is probably not resonant, it is essential that it be a time-dependent process acting on the non-equilibrium state of the exciton-rich absorber, allowing it to pump energy into the motion of the charge carries.

The photogenerated electrons form a fluid, so that the self-oscillating piston acts as a pump that produces a hydrodynamic pressure equivalent to the chemical potential of the electrons.  In open-circuit conditions this pressure gives the open circuit voltage $V_{\rm oc}$.  The self-oscillation of the internal capacitor would stop upon saturation of the charge accumulated at the terminals to its maximum $Q_{\rm max}$ (with a corresponding saturation of the electrical work output $W_{\rm max} = Q_{\rm max} V_{\rm oc}$), because the pressure that the gas of photo-generated charge exerts on the internal capacitor plates would be unable to overcome their electrostatic interaction with the charged terminals (anode and cathode).  This is analogous to how a hydrodynamic pump connected to two perfectly leakless vessels would stop if the force driving the vanes could no longer overcome the pressure difference between the vessels.  In both cases, the inevitable leakage implies that, upon saturation, the pumping does not stop altogether, but rather proceeds weakly so as to compensate for the leaks.

For a closed circuit, this pressure corresponds to the non-conservative force driving the photocurrent's circulation.  As for an ordinary hydrodynamic pump, the oscillations of the pressure are smoothed out by the fluid's viscosity, so that far from the pumping mechanism one observes only a static pressure gradient.  The dynamical picture of the PV device as an autonomous heat engine is now complete: the piston (an internal capacitor with a degree of freedom corresponding to a plasmonic or vibronic oscillatory mode) modulates the state of the working substance (the conducting electrons) in such a way that it absorbs more heat from the hot bath (incoming photons) than it rejects into the cold bath (thermal phonons at ambient temperature).  This energy is available to sustain the oscillation of the PV piston against the losses due to internal damping and the external load.  Part of the mechanical work in the piston's oscillation is converted into electrical work by separating the charges and moving them against the macroscopic electric field between the device's terminals.

\section{Outlook}
\la{sec:outlook}

The picture of dynamic charge separation that we have described predicts several phenomena that are absent in the traditional quasi-static description of PV devices, namely
 
\begin{itemize}
	\item (A) charge separation should exhibit oscillatory features consistent with the piston dynamics,
	\item (B) an illuminated PV device should emit weak radiation at the piston's frequency, and
	\item (C) excitation of the solar cell with external coherent radiation at the piston's frequency should enhance the efficiency of charge separation and the magnitude of the photocurrent.
\end{itemize}

Experimental confirmation of (A) \cite{Falke2014, DeSio, Gelinas2014} and (C) \cite{Bakulin2012,Bakulin2015} has already been provided in OPV devices, as summarized in \Sec{sec:pheno}. Theoretical work further supports the need to include dynamics in the description of charge separation \cite{Tamura2013}. The recent observation of THz emission from perovskite lattice \cite{Guzelturk2018} provide a strong support for (B), in particular because the applied illumination was equivalent to 1-sun (i.e. standard operating conditions).  Until now, however, all of these phenomena have been generally interpreted as specific to organic (or perovskite) materials, rather than as the general and fundamental mechanism of charge separation that accounts for the photovoltaic effect.

Combining our model with existing experimental evidence from the field, we conclude that strong coupling between electronic states and specific vibrational modes probably provides the necessary piston-like motion to achieve charge separation in OPVs. Specifically, for OPVs this suggests strong coupling between the electronic states of the $\pi$-electrons and the vibrational modes of the conjugated molecular backbone. An example of this is the observation of coherent oscillations of the C=C modes of the donor polymer and the pentagonal pinch mode of the acceptor fullerene during charge transfer in a polymer-fullerene blend \cite{Falke2014}. In the case of strong coupling, the piston's damping coefficient $\gamma$ in \Eq{eq:x} is proportional to the line width of the vibrational mode. Additional coupling to other modes increases damping, which is a loss mechanism, particularly if it leads to over-damping of the piston's motion. An interesting example is provided in the study by Bakulin et al.\cite{Bakulin2015}, where the authors demonstrate enhancement and suppression of the photocurrent in OPV devices via selective excitation of molecular vibrational modes.

Another consequence of the dynamical picture of energy conversion in photovoltaic devices is that the total efficiency will be limited by the pure loss associated with the linear dissipation $\gamma$ in \Eq{eq:x}.  If we estimate that most of the energy $E_{\rm in}$ injected into the piston's oscillation by the pressure of the excitons is transferred to the pumping of the separated charges during one half of the cycle (corresponding to the expansion of the internal capacitor), we may estimate the electrical work output $W$ per cycle as
\be
W = E_{\rm in} e^{-\gamma \tau / 2}
\ee
where $\tau \equiv \omega / 2 \pi$ is the period of the oscillation.  The overall efficiency of the solar cell is therefore reduced by a factor of
\be
\eta_{\rm piston} \equiv \frac{W}{E_{\rm in}} = e^{-\pi \gamma / \omega} = e^{-\pi/Q}
\ee
where $Q \equiv \omega / \gamma$ is the quality factor of the corresponding oscillation when resonantly driven.

This quality factor has not been directly measured in solid-state cells, but observations of plasma oscillations in bulk semi-conductors show relatively low $Q$'s (see, e.g., \cite{plasma2}).  This might explain why even the best solid-state cells fall short of the SQ efficiency by a factor of about 0.8 \cite{Miller2012}.  Infrared spectroscopy of the dominant molecular oscillations associated with charge separation in OPVs show significantly higher quality factors, perhaps of order 70 (see, e.g., \cite{DeSio}) corresponding to $\eta_{\rm piston} = 0.95$).  We therefore expect that the efficiency of OPVs is currently being limited primarily by the bandgap not being optimized to the solar spectrum \cite{Kroon2008}, as well by the consumption of part of the work extracted by the piston in the ionization of the CTE.

Our dynamical model thus explains why OPV systems with low driving energy $\Delta E$ may still exhibit efficient charge separation \cite{Liu2016, Kurpiers2018}. Interestingly, this opens up the possibility of a dedicated design of OPV systems that {\it independently} maximizes both the photocurrent (via efficient charge separation) and the photovoltage (via tuning of the molecular frontier orbitals).  The dynamical picture of charge separation is general and can be extended beyond OPV to explain the physics of other PV technologies. Ultimately, it implies that charge separation and current flow in PV devices cannot be understood as a spontaneous process driven only by photon energy, but must be explained by introducing engine-like dynamics internal to the solar cell.  We should, therefore, re-examine the quasi-static picture of silicon PV \cite{cycle} that has been ubiquitously adapted to describe the functioning principles of all PV technologies.

As we have emphasized here, extensive research on nanoscale transport has established that the two necessary conditions for pumping of charge are a symmetry breaking (which gives the direction of the pumping) and a time periodicity (i.e., a nonequilibrium forcing), which we have argued must come from a self-oscillating piston.  These conditions may be present in PV devices that do not contain an internal homojunction or heterojunction.  Our dynamical picture can therefore help to answer longstanding questions about what drives charge separation in high efficiency thin film PV devices, such as perovskite solar cells \cite{Uratani2017, Guzelturk2018}.  Furthermore, this picture may unravel the mystery of the anomalous bulk PV effect observed in ferroelectric semiconductors \cite{Spanier2016} and in semiconductors subject to mechanical strain \cite{Yang2018}.  It is also worth recalling that the PV effect was first reported in 1839 in an electrochemical system \cite{Becquerel} and that electrochemical self-oscillations have been reported and studied in recent decades (see, e.g. \cite{electrochem1,electrochem2}).

The concept of dynamic PV energy conversion may also be of considerable practical value. The quasi-static description of the solar cell does not provide any information about the power that can be obtained from a PV device \cite{Markvart}.  A more detailed modeling of the internal piston's dynamics could guide the optimization of the power generated by new PV devices.   Moreover, a fully dynamical description of work extraction for the broad class of energy transducers that can be characterized as cyclic engines ---including PV cells, thermoelectric generators \cite{thermoelectric-RA}, fuel cells \cite{fuelcells-RA}, biological engines, and even meteorological engines such as hurricanes and ocean waves \cite{instabilities}--- is of interest across many areas of pure and applied science, and in both quantum and classical physics.  We hope that such a perspective will open up new dimensions in the drive towards smart design of efficient new technologies for energy conversion. \\

{\bf Acknowledgements:} We thank Bob Jaffe, Sander Mann, Tom Markvart, and Rinke Wijngaarden for useful discussions.  Some of this work was completed while RA visited the Kavli Institute for Theoretical Physics, in Santa Barbara, with support from the US National Science Foundation (grant no.\ NSF PHY-1748958).  DG-K's work was supported by the Center for Excitonics, an Energy Frontier Research Center funded by the US Department of Energy under award DE-SC0001088 (Solar energy conversion process), and later by the Gordon and Betty Moore Foundation as a Physics of Living Systems Fellow (grant GBMF4513).  AJ's work was supported by the University of Costa Rica's Vice-rectorate for Research (project no.\ 112-B6-509) and by the European Union's Horizon 2020 research and innovation program under the Marie Sk{\l}odowska-Curie grant agreement no.\ 690575.  EvH's work was supported by an ECHO grant from The Netherlands Organisation for Scientific Research (NWO), project no.\ 712.017.001.


\end{document}